\documentclass[pre,aps,showpacs,preprintnumbers]{revtex4}

\usepackage{graphicx}% Include figure files
\usepackage{dcolumn}% Align table columns on decimal point
\usepackage{bm}% bold math
\usepackage{amssymb}
\usepackage{pifont}
\usepackage{amsmath}
\usepackage{times}
\usepackage[nooneline]{subfigure}
\usepackage{ifthen}
\usepackage{xcolor}

\newcommand\hxt{h(\mathbf{x},t)}
\newcommand\hxjt{h(\mathbf{x}_\mathbf{j},t)}
\newcommand\hxet[2]{
  \ifthenelse{#1=1}{h(\mathbf{x}+\mathbf{e}_{#2}\Delta x,t)}{%
    \ifthenelse{#1=-1}{h(\mathbf{x}-\mathbf{e}_{#2}\Delta x,t)}
      {h(\mathbf{x}-{#1}\mathbf{e}_{#2}\Delta x,t)}}
}
\newcommand\fop[1]{\Delta^{\text{f}}_{#1}}
\newcommand\bop[1]{\Delta^{\text{b}}_{#1}}
\newcommand\xb{\mathbf{x}}
\newcommand\kb{\mathbf{k}}
\newcommand\qb{\mathbf{q}}
\newcommand\xj{\mathbf{x}_\mathbf{j}}
\newcommand\xjp{\mathbf{x}_\mathbf{j}'}
\newcommand\eiqx{e^{i\mathbf{q}\cdot\mathbf{x}}}
\newcommand\emiqx{e^{-i\mathbf{q}\cdot\mathbf{x}}}
\newcommand\leqs{\leqslant}

\begin{document}

\title{Pseudospectral versus finite-differences schemes in the numerical
integration of stochastic models of surface growth} 

\author{Rafael Gallego} \email[]{rgallego@uniovi.es} 

\affiliation{Departamento de Matem\'aticas, Universidad de Oviedo, Campus de
Viesques, E-33203 Gij\'on, Spain} 

\author{Mario Castro}\email[]{marioc@upcomillas.es} 

\affiliation{Grupo Interdisciplinar de Sistemas Complejos (GISC) and Grupo de
Din\'amica No Lineal (DNL), Escuela T\'ecnica Superior de Ingenier{\'\i}a
(ICAI), Universidad Pontificia Comillas, E-28015 Madrid, Spain} 

\author{Juan M. L\'opez}\email[]{lopez@ifca.unican.es} 

\affiliation{Instituto de F{\'\i}sica de Cantabria (IFCA), CSIC--UC, E-39005
Santander, Spain}

\date{\today}

\begin{abstract}
We present a comparison between finite differences schemes and a pseudospectral
method applied to the numerical integration of stochastic partial differential
equations that model surface growth. We have studied, in 1+1 dimensions, the
Kardar, Parisi and Zhang model (KPZ) and the Lai, Das Sarma and Villain model
(LDV). The pseudospectral method appears to be the most stable for a given time
step for both models. This means that the time up to which we can follow the
temporal evolution of a given system is larger for the pseudospectral method.
Moreover, for the KPZ model, a pseudospectral scheme gives results closer to the
predictions of the continuum model than those obtained through finite difference
methods. On the other hand, some numerical instabilities appearing with finite
difference methods for the LDV model are absent when a pseudospectral
integration is performed. These numerical instabilities give rise to an
approximate multiscaling observed in the numerical simulations. With the
pseudospectral approach no multiscaling is seen in agreement with the continuum
model.
\end{abstract}


\maketitle

\section{Introduction}
Kinetic surface roughening of surfaces growing in nonequilibrium conditions has
been intensively studied for the last two decades \cite{Barabasi, Krug-rev,
Pimpi}. Theoretical approaches make use of both discrete atomistic simulations
and stochastic continuum equations for the evolution of the coarse-grained
surface height $h(\mathbf{x},t)$. There is overwhelming experimental evidence
that surfaces under general nonequilibrium growth conditions can develop
scale-invariant correlations in space and time, which supports the hope of a
unified theoretical framework to understand kinetic roughening phenomena
from first principles. The aim is at identifying the various dynamical
universalities of growth associated with different sets of symmetries and/or
conservation laws. It is believed that only these basic elements largely
determine the universality class and the value of the
corresponding critical exponents.  In theoretical studies attention is therefore
focused on symmetries and only the most relevant terms (in the renormalization
group sense) are expected to be required to describe a particular class of
growth.

Universality classes of growth are generically represented by stochastic partial
differential equations,
\begin{equation}
\label{langevin} \partial_t h = {\cal G}(\nabla h) + \eta({\bf
x},t),
\end{equation}
where $h({\bf x},t)$ is the height of the interface at substrate position ${\bf
x}$ and time $t$. The external noise $\eta({\bf x},t)$ represents the influx of
atoms on the surface. The function ${\cal G}(\nabla h)$ defines a particular
model and incorporates the relevant symmetries and conservation laws. In
particular, invariance under translation along the growth and substrate
directions as well as invariance in the election of the time origin rule out an
explicit dependence of ${\cal G}$ on $h$, ${\bf x}$ and $t$. Very often the
presence of nonlinearities in ${\cal G}$ require the use of perturbative
renormalization techniques to obtain analytical approximations for the critical exponents,
which can then be compared with Monte Carlo simulations of atomistic models and
experiments. A perturbative renormalization approach invariably provides the
critical exponents as a series expansion on the parameter $\epsilon = d_c - d$,
where the critical dimension $d_c$ can be very high when compared with the
dimensions of physical interest (usually $d=1$ or $2$). Only in a few lucky
cases some extra symmetries produce cancellation of higher order loop diagrams
that results in a scaling relation between exponents to be exact to all orders
in the perturbative expansion. More often than not, the case is that we only
have approximations to the critical exponents valid up to a certain order in
$\epsilon$ and a great deal of elaborated algebraic effort is required to
improve our approximation up to the next order. This often makes direct
numerical integration of Eq.\ \eqref{langevin} an extremely useful and necessary
tool as the most reliable source of precise values for the critical exponents.

Numerical schemes to integrate continuum surface growth equations like
Eq.\ \eqref{langevin} in $1+1$ and $2+1$ dimensions tend to be unsophisticated. In
most cases a straightforward finite-differences (FD) method on a lattice does an
excellent job and provides highly precise values for the critical exponents, at
least in dimensions of experimental relevance. In this approach (see details
below) one basically approximates the continuous height field, $h(\mathbf{x},t)$, by its
values on the lattice sites, $h_{i}(t)$, and derivatives by differences between
neighboring sites. More clever choices of the discretization rule have been
shown to be useful to obtain better agreement with exact properties of the
continuum solutions \cite{Lam}, which could not be obtained by using a
conventional discretization, like the nominal values of the continuum equation
parameters.

However, the use of FD schemes sometimes poses some important problems
\cite{Moser, Doherty, Tu}. In particular Dasgupta {\it et. al.} have shown
\cite{Dasgupta, Dasgupta2} by means of numerical simulations that discretized
versions of commonly studied nonlinear growth equations exhibit a instability in
the sense that single pillars (grooves) become unstable when their height
(depth) exceeds a critical value. In some cases these instabilities are not
present in the corresponding continuum equations, indicating that the behavior
of the discretized versions is indeed different from their continuum
counterparts. It is important to remark that this
pillar/groove instability is actually generic to the FD discretizations of a
large class of nonlinear growth equations, including the Kardar-Parisi-Zhang
(KPZ) \cite{Kardar} or the Lai-Das Sarma-Villain (LDV) equations
\cite{Lai-DasSarma, Villain,Wolf-Villain}. This is a puzzling result because the
corresponding continuum equations are {\em not} unstable.

In many situations, like for instance in KPZ, the existence of this instability is of
little significance for practical purposes and one can actually carry out a correct
numerical integration by using FD schemes. The reason is that one is mostly interested in
the growth from a flat (or almost) surface initial condition and common relaxation
mechanisms do not favor the formation of large pillars or grooves. In these cases the
instability is only realized if the initial condition is prepared in such a state that
there is a pillar/groove of size above the threshold on a otherwise flat surface, which is
highly artificial and usually uninteresting for practical purposes. However, as already
pointed out in Ref.\ \cite{Dasgupta2}, there is a large class of systems for which the
instability of any FD scheme is inevitable. Specifically, discrete versions of models
exhibiting anomalous kinetic roughening \cite{Krug, DasSarma, Lopez97, Lopez99, Ramasco,
Lopez05, Bhatta} will certainly show this kind of instability at sufficiently long times.
The reason being that anomalous scaling is associated with a nontrivial dynamics of the
average surface gradient (local slope), so that $\langle \overline{(\nabla
h)^2}\rangle^{1/2} \sim t^{\kappa}$, with $\kappa > 0$ \cite{Krug, Lopez99}. Therefore,
systems exhibiting anomalous roughening will dynamically generate large local height
differences, no matter how flat the initial condition is. As a consequence, provided that
a simulation is run long enough, the surface will produce pillars/grooves above the
critical value for the instability to appear, like for instance is the case for LDV. 

One of the models we study in this paper is 
the LDV equation \cite{Lai-DasSarma, Villain, Wolf-Villain}
\begin{equation}
\label{eqLDV0}
  \partial_th(\mathbf{x},t)= - K \nabla^4h+\lambda\nabla^2(\nabla h)^2+
  \xi(\mathbf{x},t),
\end{equation}
where the noise is Gaussian distributed and delta correlated,
\begin{equation}
 \langle\xi(\mathbf{x},t)\xi(\mathbf{x}',t')\rangle=2D
\delta(\mathbf{x}-\mathbf{x}')\delta(t-t').
\end{equation}
This model constitutes a minimal model for the long wavelength behavior of
surface growth under ideal molecular-beam epitaxy conditions. The LDV model is
interesting in many respects and has been the focus of a lot of attention in the
literature \cite{Lai-DasSarma, Villain, Wolf-Villain, Lopez99, Bhatta,
Krug_Plisch, Kim}. 

Numerical simulations of discrete versions of \eqref{eqLDV0} in $1+1$ dimensions
have reported \cite{Dasgupta2} a finite, albeit small, anomalous exponent
$\kappa \approx 0.08$, possibly indicating a logarithmic dependence. A
theoretical prediction \cite{Lopez99} based on Flory-type arguments predicted
$\kappa \approx 1/11$ (see however \cite{note}). Therefore, from the discussion
above, one would expect a discrete version of \eqref{eqLDV0} to become unstable.
This problem was studied by Dasgupta {\it et. al.} and they showed
\cite{Dasgupta, Dasgupta2} that FD algorithms were actually
unstable at long times. They also estimated the critical height step to be
around $h_c(\lambda) \approx A/\lambda$ with $A \approx 20$ for
\eqref{eqLDV0} with $K=D=1$, which clearly shows that the instability will appear the
sooner the larger the nonlinear coefficient is. Those authors also found that the
addition of higher-order nonlinearities in the FD version of the model controls
the numerical instabilities and renders a stable surface, but with intermittent
fluctuations and multiscaling properties of surface correlations. It has been
claimed that higher-order nonlinearities of the form $\lambda_{2n} \nabla^2
(\nabla h)^{2n}$, with $n>1$, may play an important role in LDV universality
class because they are infinitely many marginally relevant nonlinear terms
\cite{Lai-DasSarma, Tu, Krug, Bhatta}.

These results can be compared with FD integration schemes for the KPZ equation
\begin{equation}\label{eqKPZ0}
  \partial_th(\mathbf{x},t) = \nu\nabla^2 h + \lambda (\nabla h)^2+
  \xi(\mathbf{x},t).
\end{equation}

It has been shown \cite{Dasgupta, Dasgupta2} that discrete versions of Eq.\
\eqref{eqKPZ0} were stable, unless isolated grooves of large enough size are
included in the initial state. The reason being that KPZ exhibits conventional
(no anomalous) scaling and local slopes are thus rapidly converging towards a
constant. Under general conditions the constant is much smaller than the
critical slope $h_c$ for the KPZ discretization to become unstable and so, large
slopes are not spontaneously generated by the dynamics. 

In reference~\cite{Giada2} a study of the 1D and 2D KPZ equation using
finite-differences and pseudospectral integration methods is presented.  Authors
claim that a pseudospectral method gives results closer to the continuum limit
than finite-differences methods. They show how a pseudospectral method
reproduces the exact value of the global width of the steady state interface
within error bars whereas a finite-differences method with conventional
discretization of the nonlinear term entails significant differences in the
amplitude value. They also use pseudospectral computations to reproduce the most
reliable values of the dynamical exponents obtained through discrete growth
models.

In this paper we discuss the validity of FD integration algorithms in the
presence of anomalous roughening. We compare the accuracy, stability and overall
performance of FD methods versus pseudospectral (PS) schemes applied to the
paradigmatic examples of the KPZ and LDV equations. We claim that the
instability previously found in FD discretizations is spurious and non physical,
therefore, FD should be generally avoided in numerical simulations of continuum
growth models with anomalous scaling. We argue that the main reason for the
adequacy of PS to attack growth problems with anomalous scaling is that spatial
derivatives are {\em more accurate} than in FD methods, where one implicitly
assumes that the step height is small. Our conclusions are based on numerical
analysis of KPZ and LDV equations in $1+1$ dimensions by means of FD and PS
integration schemes. Our results when comparing both techniques are conclusive:
{\it (i)} PS methods are stable against isolated pillars/grooves, while FD are
not, {\it (ii)} under the same conditions PS schemes take much longer than FD to
get to a numerical overflow, and {\it (iii)} PS schemes give well behaved
correlation functions with no trace of multiscaling.  Finally, we will discuss
the implications of our results for the appearance of multiscaling in discrete
models proved to be in the same universality class as LDV
equations~\cite{Vvedensky}.

%%%%%%%%%%%%%%%%%%%%%%%%%%%%%%%%%%%%%%%
\section{Numerical integration schemes}
%%%%%%%%%%%%%%%%%%%%%%%%%%%%%%%%%%%%%%%
In order to perform a numerical integration of Eqs.\ \eqref{eqKPZ0} and
\eqref{eqLDV0} the parameters can easily be rescaled to have only one
independent control parameter-- namely, the coupling constant. As it is
customary, one can work with dimensionless variables $h$, $\mathbf{x}$ and $t$
so that all parameters but one are set to unity, so we have 
\begin{equation}\label{eqKPZ}
  \partial_th(\mathbf{x},t)=\nabla^2h + g (\nabla h)^2+\eta(\mathbf{x},t),
\end{equation}
for the KPZ equation, where the dimensionless coupling constant is $g = \lambda
\sqrt{2D/\nu^3}$. We can also write the LDV equation in dimensionless form
\begin{equation}\label{eqLDV}
  \partial_th(\mathbf{x},t)=-\nabla^4h+g\nabla^2(\nabla h)^2+\eta(\mathbf{x},t),
\end{equation}
where the coupling constant is $g = \lambda \sqrt{2D/K^3}$
and $\eta(\mathbf{x},t)$ is a
Gaussian noise with mean zero, unit variance and correlations
$\langle\eta(\mathbf{x},t) \eta(\mathbf{x}',t') \rangle =
\delta(\mathbf{x}-\mathbf{ x}') \delta(t-t')$.

Let us now summarize the idea behind FD and PS integration schemes and introduce some
useful definitions. Equations \eqref{eqKPZ} and \eqref{eqLDV} can be cast in the form
\begin{equation}\label{eq:generalModel}
  \partial_t h(\mathbf{x},t)=\mathcal{L}[h](\mathbf{x},t)+
  \Phi[h](\mathbf{x},t)+\eta(\mathbf{x},t),
\end{equation}
where $\mathcal{L}[h]$ is a linear functional of $h$ and $\Phi[h]$ is
another functional containing the nonlinear terms. 

%======================================
\subsection{Finite-differences methods}
%======================================

We consider a $d$-dimensional lattice with periodic boundary conditions with 
uniform spacing $\Delta x$ in each direction. The positions of the nodes in the
lattice are given by
\begin{equation}
  \xj=\Delta x(j_1,j_2,\dotsc,j_d),\ 0\leqs j_i\leqs N_i-1,\ 1\leqs i\leqs d.
\end{equation}
where $N_i$ is the lattice size in the $i$-th direction. Using a one step
Euler's method to compute the temporal derivative, the evolution of a system
governed by Eq.~\eqref{eq:generalModel} is given by:
\begin{equation}\label{eq:FD}
  h(\xj,t+\Delta t)=
  h(\xj,t)+\Delta t\bigl(
  \mathcal{L}[h](\xj,t)+
  \Phi[h](\xj,t)\bigr)+\sqrt{\dfrac{\Delta t}{(\Delta
  x)^d}}\,\eta(\xj,t).
\end{equation}
where $\Delta t$ is the time step and the stochastic variables $\eta(\xj,t)$
have zero mean and correlations $\langle\eta(\xj,t) \eta(\xjp,t')\rangle =
\delta_{\mathbf{j},\mathbf{j}'}\delta(t-t')$. We took the $\eta$ variables as
Gaussian random numbers (other distributions can be used as long as they satisfy
the central limit theorem). 

In finite difference methods, derivatives are computed by truncating the Taylor
series of the field up to certain order. Let us introduce the finite difference
operators $\fop{j}$ and $\bop{j}$ which are, respectively, the forward and
backward difference operators along the $j$ direction:
\begin{align}
  \fop{j}\hxt &= h(\mathbf{x}+\mathbf{e}_j\Delta x)-\hxt,\label{eq:opf}\\
  \bop{j}\hxt &= \hxt - h(\mathbf{x}-\mathbf{e}_j\Delta x).\label{eq:opb}
\end{align}
In terms of these operators, the linear parts of Eqs. \eqref{eqKPZ} and
\eqref{eqLDV} are, up to second order of approximation, given by:
\begin{gather*}
  \mathcal L_{\text{KPZ}}(\xj,t)=(\nabla^2 h)(\xj,t) =
  (\Delta x)^{-2}\sum_{i=1}^d\fop{i}\bop{i}\hxjt,\\
  \mathcal L_{\text{LDV}}(\xj,t)=-(\nabla^4 h)(\xj,t) =-\nabla^2(\nabla^2 h)=
  (\Delta x)^{-4}\sum_{i,j=1}^d\fop{i}\bop{i}\fop{j}\bop{j}\hxjt.
\end{gather*}
The explicit expressions in $1+1$ dimensions are
\begin{gather*}
  \mathcal L_\text{KPZ}[h]=(\Delta
  x)^{-2}(h_{i+1}-2h_i+h_{i-1}),\\
  \mathcal L_\text{LDV}[h]=-(\Delta x)^{-4}
  (h_{i+2}-4h_{i+1}+6h_i-4h_{i-1}+h_{i-2}),
\end{gather*}
where $x_i=i\Delta x$, $i=0,\dotsc,N-1$ are the positions of the nodes in the
lattice and $h_i=h(x_i,t)$. Regarding the nonlinear terms we consider for the
gradient square the usual symmetric discretization: 
\begin{equation*}
  (\nabla h)^2(\mathbf{x},t)=\frac{1}{2}(\Delta x)^{-2}
  \sum_{i=1}^d[(\fop{i}+\bop{i})\hxt]^2
\end{equation*}
that in $1+1$ dimensions becomes
\begin{equation}\label{eq:usualDiscKPZ}
  (\nabla h)^2(x_i,t)=\frac{1}{2}(\Delta x)^{-2}
  (h_{i+1}-h_{i-1})^2.
\end{equation}

In the case of the KPZ equation, other discretizations of the nonlinear term
have been proposed~\cite{Lam,Giacometti}. We mention the Lam and Shin
discretization (LS)~\cite{Lam}
\begin{equation*}
  (\nabla h)^2(\xj,t)=
  \frac{1}{3}(\Delta x)^{-2}\sum_{i=1}^d
  \bigl\{[(\fop{i}+\bop{i})\hxjt]^2-\bigl(\fop{i}\hxjt\bigr)
  \bigl(\bop{i}\hxjt\bigr)\bigr\},
\end{equation*}
so that in 1+1 dimensions we have
\begin{equation}\label{eq:LSDiscKPZ}
  (\nabla h)^2(x_i,t)=\frac{1}{3}(\Delta x)^{-2}
  [(h_{i+1}-h_i)^2+(h_{i+1}-h_i)(h_{i}-h_{i-1})+
  (h_{i}-h_{i-1})^2].
\end{equation}
LS discretization has two interesting features in 1+1 dimensions: {\it (i)}
the effective parameter $g$ agrees with its nominal value, and {\it (ii)} the
probability
distribution of the discretized version in the steady state can be computed
exactly and it turns out to be the probability distribution of the continuum
equation for all values of $g$. It has been argued~\cite{Lam} that this
discretization allows to recover some results predicted by the continuum model
while discrepancies when using the conventional discretization
\eqref{eq:usualDiscKPZ} have been observed~\cite{Lam}. 

In the following we use a lattice spacing $\Delta x = 1$. As is customary in
this kind of simulations, hydrodynamic limit is achieved by increasing the
number of lattice sites $N$. In numerical integrations of continuous growth
models one avoids to perform the $\Delta x \to 0$ limit with fixed $L$, which
would lead the system towards the {\em linear} critical point, since the
coupling constant of the discretized equation is $g \to 0$ as $\Delta x \to 0$.
A fixed lattice spacing $\Delta x$ in the limit $L \to \infty$ is always
preferred as it best drives the system towards the nontrivial critical point.

%===================================
\subsection{Pseudospectral method}
%===================================

To compare with FD methods we have considered a numerical scheme consisting of a
spectral method in space together with a Euler's method in time. We assume that
the field $\hxt$ satisfies periodic boundary conditions in the multidimensional
interval $[0,L]^d$ and we represent it as a truncated Fourier series
\begin{equation*}
  h_N(\xb,t)=\sum_{\kb\in\Gamma_N}\tilde h_\kb(t)\eiqx,\quad 
  \qb=\frac{2\pi}{L}\kb.
\end{equation*}
The set $\Gamma_N$ over which the sum is taken is given by
$\Gamma_N=\{(k_1,k_2,\dotsc,k_d)\;\diagup\;-N/2\leqs k_i\leqs N/2-1, 1\leqs
i\leqs d\}$, and the \mbox{$\tilde h_\kb(t)$'s} are the Fourier coefficients of
$h$, defined as
\begin{equation*}
  \tilde h_\kb(t)=\frac{1}{L^d}\int_{[0,L]^d} d\xb\,\hxt\emiqx.
\end{equation*}
The noise term $\eta$ is also replaced by its expansion $\eta_N$ in Fourier
modes. When $N\rightarrow\infty$ the usual Fourier series is recovered. On the
other hand, when $h$ and $\eta$ are replaced by $h_N$ and $\eta_N$ respectively
in Eq. \eqref{eq:generalModel}, the residual
\begin{equation*}
  R_N(\mathbf{x},t)=\partial_t h_N-\mathcal L[h_N]-\Phi[h_N]-\eta_N  
\end{equation*}
will be not null in general. By requiring $R_N$ to be orthogonal to the
functions $\{\eiqx, \kb\in\Gamma_N\}$, we obtain a set of ODEs for the Fourier
coefficients of $h$. This procedure is actually equivalent to project the
equation onto a subspace of orthogonal polynomials of degree $\leqs N/2$. Then,
by imposing the orthogonality condition
\begin{equation*}
  \int_{[0,L]^d} d\xb\,R_N(\mathbf{x},t)\emiqx=0,\quad
  \kb\in\Gamma_N
\end{equation*}
we obtain
\begin{equation}\label{eq:pseGeneral}
  \frac{d\tilde h_\kb(t)}{dt}=
  \omega_\kb\tilde h_\kb(t)+\tilde\Phi_\kb(t)+\tilde\eta_\kb(t),\quad
  \kb\in\Gamma_N.      
\end{equation}
The quantity $\omega_\kb$ is the linear dispersion relation, which is obtained
through the Fourier transform of the linear part of the equation; it is
$\omega_\kb=-\qb^2$ for Eq. \eqref{eqKPZ} and $\omega_\kb=-\qb^4$ for Eq.
\eqref{eqLDV}. The $\tilde\Phi_\kb(t)$'s are the Fourier coefficients of the
nonlinear terms and are given by the following convolution sums:
\begin{equation*}
  \tilde\Phi_\kb(t)=
  \begin{cases}
    -g\displaystyle\sum_{\kb_1+\kb_2=\kb}\qb_1\cdot\qb_2\,\,
      \tilde h_{\kb_1}\,\tilde h_{\kb_2} \quad &\text{(KPZ)},\\[0.3cm]
    g\qb^2\displaystyle\sum_{\kb_1+\kb_2=\kb}\qb_1\cdot\qb_2\,\,
      \tilde h_{\kb_1}\,\tilde h_{\kb_2}  &\text{(LDV)}.
  \end{cases}
\end{equation*}
Regarding the Fourier coefficients of the noise, it is easy to verify that
$\tilde\eta_\kb(t)$ are complex Gaussian variables with zero mean and
correlations
\begin{equation*}
  \langle\tilde\eta_\kb(t)\tilde\eta_{\kb'}(t')\rangle=
  \frac{1}{L^d}\delta_{\kb,-\kb'}\delta(t-t').
\end{equation*}

The Fourier coefficients $\tilde h_\kb$ are in general difficult to compute. In
addition, even the simplest nonlinearities make it computationally expensive the
task of going from real space to Fourier space and viceversa. For these
reasons, we consider a discretized space with $N$ nodes in each direction
\begin{equation*}
  \xj=\frac{L}{N}(j_1,j_2,\dotsc,j_d),\quad
  0\leqs j_i\leqs N-1,\quad 1\leqs i\leqs d,
\end{equation*}
and we use the discrete Fourier transform $\mathcal F$ to
integrate~\eqref{eq:pseGeneral}. The discrete Fourier coefficients depend only
on the values of the field at the nodes $\xj$ and are given by (direct discrete
Fourier transform):
\begin{equation*}
  \hat h_\kb=\mathcal{F}[h_\mathbf{j}]=\frac{1}{N^d}
  \sum_{\mathbf{j}}h_\mathbf{j}(t)
  e^{-i\mathbf{q}\cdot\xj}
\end{equation*}
where $h_\mathbf{j}(t)=h(\xj,t)$. We have the inversion formula
(inverse discrete Fourier transform):
\begin{equation*}
  h_\mathbf{j}(t)=\mathcal{F}^{-1}[\hat h_\kb]=\sum_{\kb\in\Gamma_N}\hat h_\kb 
  e^{i\mathbf{q}\cdot\xj}.
\end{equation*}
Then, we replace the continuum Fourier coefficients in \eqref{eq:pseGeneral} by
the discrete ones, so that
\begin{equation}\label{eq:pseGeneral2}
  \frac{d\hat h_\kb(t)}{dt}=
  \omega_\kb\hat h_\kb(t)+\hat\Phi_\kb(t)+\hat\eta_\kb(t).
\end{equation}
Note that Eq.~\eqref{eq:pseGeneral2} is now written in terms of the discrete
Fourier coefficients $\hat h_\kb$. To integrate \eqref{eq:pseGeneral2} we
perform the following change of variables based on the solution of the linear
equation:
\begin{equation*}
  \hat h_\kb(t)=e^{\omega_\kb t}\hat z_\kb(t)+\hat R_\kb(t)
\end{equation*}
where
\begin{equation*}
  \hat R_\kb(t)=e^{\omega_\kb t}\int_0^t du\,e^{-\omega_\kb u}\hat\eta_\kb(u).
\end{equation*}
The $\hat z_\kb$'s satisfy the equations
\begin{equation}\label{eq:zetaODE}
  \frac{d\hat z_\kb(t)}{dt}=\hat\Phi_\kb(t)e^{-\omega_\kb t}.
\end{equation}
The set of ODEs \eqref{eq:zetaODE} can be solved by using one of the several
algorithms available for stochastic differential equations (Euler, Runge-Kutta,
predictor-corrector methods, etc.). Considering a one step Euler's method to
integrate \eqref{eq:zetaODE} and going back to the original variable $\hat
h_\kb$, we are finally left with
\begin{equation}\label{eq:intFactor}
  \hat h_\kb(t+\Delta t)=e^{\omega_\kb\Delta t}
  [\hat h_\kb(t)+\Delta t\,\hat\Phi_\kb(t)]+\hat r_\kb(t).
\end{equation}
Equation~\eqref{eq:intFactor} can be reinterpreted as an Euler scheme with
time-step (the factor multiplying the nonlinear term) $e^{\omega_\kb\Delta
t}\Delta t$, so our algorithm provides a smaller time-step for the smallest
length-scales (so it is intrinsically multiscale). This represents a
significant improvement with respect to the pseudospectral method used
in~\cite{Giada2} which is just Eq.~\eqref{eq:pseGeneral2} integrated with a one
step Euler's method.

Assuming that $\omega_\kb=\omega_{-\kb}$, the variables $\hat
r_\kb(t)$ can be obtained as
\begin{equation*}
  \hat r_\kb(t)=
    \sqrt{\frac{e^{2\omega_\kb \Delta t}-1}{2\omega_\kb}\frac{1}{(\Delta
    x)^d}}\,\,\hat v_\kb(t)
\end{equation*}
where $\Delta x=L/N$ and $\hat v_\kb(t)$ are the discrete Fourier transform of a
set of Gaussian random numbers of zero mean and unit variance. Note that,
as expected, when $\Delta t\rightarrow 0$ we recover the last term on the right side
of Eq.~\eqref{eq:FD}.

The computation of the nonlinear terms for the KPZ and LDV equations in Fourier
space involves the Fourier transform of the product of two functions (actually,
the square of $\nabla h$). In general, calling these two functions $\sigma$ and $\rho$, we
need to calculate the convolution sum
\begin{equation}\label{eq:convolutionSum}
  \hat
   \Phi_\kb=
  \sum_{\begin{subarray}{c}\kb_1+\kb_2=\kb\\\kb_1,\kb_2\in\Gamma_N\end{subarray}}
  \hat \sigma_{\kb_1}\hat \rho_{\kb_2}.
\end{equation}
In one dimension, this convolution sum implies $O(N^2)$ operations, which is
computationally more expensive than a finite difference method, for which only $O(N)$
operations are needed. To speed up the computation, we used a {\em pseudospectral}
transform method to compute the Fourier transform of the nonlinear term. Starting from
$\hat \sigma_\kb$ and $\hat \rho_\kb$, the inverse transformation is used to obtain
$\sigma$ and $\rho$ in real space. Then $\sigma$ and $\rho$ are multiplied to obtain
$\Phi$ in real space. Finally, the direct Fourier transform is applied to obtain the $\hat
\Phi_\kb$. In terms of the discrete Fourier operator $\mathcal F$,
this pseudospectral calculation can be written as follows:
\begin{equation*}
  \hat\Phi_\kb={\mathcal F}\bigl[\mathcal F^{-1}[\hat \sigma_\kb]\,
  \mathcal F^{-1}[\hat \rho_\kb]\bigr].
\end{equation*}
This procedure allows to evaluate the convolution sum using $O(N\log N)$
operations in one dimension. It is important to note that the Fourier
coefficients $\hat\Phi_\kb$ computed in a pseudospectral manner differ from
those obtained from a true spectral computation. The difference is the so-called
\emph{aliasing error}. For example, in one dimension, the coefficients
$\hat\Phi_k$ computed pseudospectrally turn out to be
\begin{equation*}
  \hat\Phi_k=\sum_{k_1+k_2=k}\hat \sigma_{k_1}\hat \rho_{k_2}+
  \sum_{k_1+k_2=k\pm N}\hat \sigma_{k_1}\hat \rho_{k_2}.
\end{equation*} 
The first term on the right hand side is just the convolution sum
\eqref{eq:convolutionSum} whereas the second term is the aliasing error. The
aliasing error has been proved to be asymptotically of the same order of the
error made in truncating the Fourier series. There are several recipes to remove
the aliasing. We used a well-known truncation technique usually referred to as the
3/2-rule~\cite{Canuto}.

%%%%%%%%%%%%%%%%%%%%%%%%%%%%%%%%%%%%%%%%%%%%%%%%%%%%%%
\section{Comparison of the methods in 1+1 dimensions}
%%%%%%%%%%%%%%%%%%%%%%%%%%%%%%%%%%%%%%%%%%%%%%%%%%%%%%

%========================
\subsection{Preliminaries}
%========================

\begin{figure}
\includegraphics[width=0.4\textwidth]{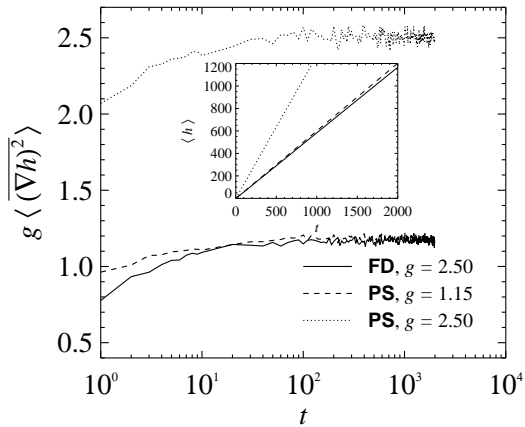}
\caption{Average over space and realizations of the nonlinear term for the 1D
KPZ equation as a function of time. The inset shows the average height of the
interface as a function of time. Here $L=128$, and the averages are taken over
100 realizations. The kind of  numerical method and the value of the nonlinear
coupling parameter for each curve are shown in the legend.\label{fig:nlinAvKPZ}}
\end{figure}

In order to compare the results provided by the FD and PS numerical methods
applied to models \eqref{eqKPZ} and \eqref{eqLDV}, we must first notice that,
for any given model and the same value of the nonlinear coupling parameter $g$,
the intensity of the nonlinear effects depends on the numerical scheme used to
integrate the equation. This fact, which has already been pointed out in
\cite{Giada}, leads to the conclusion that different algorithms cannot in
principle be compared directly. In Fig.~\ref{fig:nlinAvKPZ}, the average (both
over space and realizations) of the nonlinear term for the 1D KPZ equation is
shown in several cases. We see that, for the same value of the coupling
parameter, the PS method gives effectively a larger nonlinear term. In other
words, for the same value of $g$, the FD method underestimates the intensity of
the nonlinear term with respect to the PS method. A comparison of the two
numerical methods can be only made if the nonlinear effects are of the same
order for both of them  on average.
%================ BEGIN FIGURE ==============
%======== END FIGURE ===========

For the KPZ equation the nonlinear effects can be monitored by measuring the mean velocity
of the interface, which is given by 
\begin{equation*}
  v=\frac{g}{L}\int_0^Ldx\,\langle(\nabla h)^2\rangle.
\end{equation*}

\begin{figure}
\includegraphics[width=0.4\textwidth]{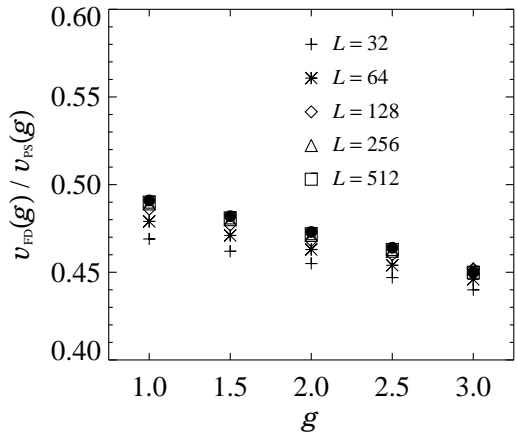}
\caption{Ratio $v_\text{FD}/v_\text{PS}$ as explained in the text as a function
of the nonlinear coupling parameter $g$ for several system
sizes.\label{fig:velRatioKPZ}}
\end{figure}

In the inset of Fig.~\ref{fig:nlinAvKPZ} the average height of the interface as
a function of time for the 1D KPZ equation is shown. The slope of this curve is
just the velocity of the interface. For the same value of $g$, the interface
obtained with the PS method moves faster. As said before, this indicates that
the nonlinear effects are stronger in the PS method than in the FD method. It is
easy to find values of $g$ such that the interface in both cases moves
approximately at the same velocity, which means that nonlinear effects are of
similar magnitude. Then, if we denote by $v_{\text{FD}}(g)$ and $v_{\text{PS}}(g)$
the mean interface velocity for the FD and PS methods, respectively, the value
of the coupling parameter $\tilde g$ such that $v_{\text{PS}}(\tilde
g)=v_{\text{FD}}(g)$ is given by
\begin{equation}\label{eq:gKPZ}
  \tilde g=g\frac{v_{\text{FD}}(g)}{v_{\text{PS}}(g)}.
\end{equation}
The ratio $v_{\text{FD}}(g)/v_{\text{PS}}(g)$ depends smoothly on both
$g$ and the system size as it can be seen in Fig.~\ref{fig:velRatioKPZ}. The
ratio of velocities of the interfaces slightly decreases with $g$ and increases
with the system size. For example, for a system size of $L=128$ and $g=2.5$,
numerically we have found that $v_{\text{FD}}(g)/v_{\text{PS}}(g)\approx 0.46$,
which means that the nonlinear effects in the PS method are approximately twice
as much stronger than those of the FD method. In Fig.~\ref{fig:nlinAvKPZ} we can
see how the nonlinear terms for both integration methods become similar when $g$
is decreased from  $2.5$ to a value of $2.5\times 0.46=1.15$ for the PS method.
%================ BEGIN FIGURE ==============
%======== END FIGURE ===========

\begin{figure}
\includegraphics[width=0.4\textwidth]{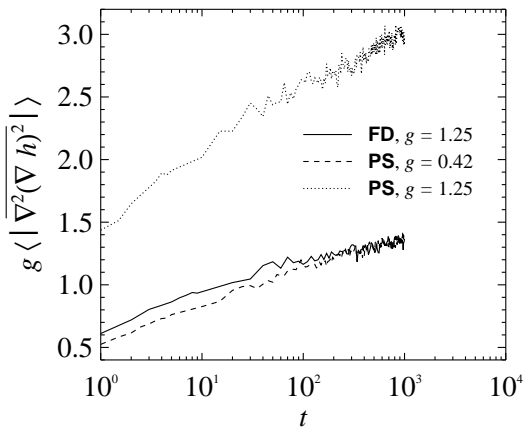}
\caption{Average over space and realizations of the absolute value of the
nonlinear term for the 1D LDV equation as a function of time. Here $L=128$, and
the averages are taken over 100 realizations. The kind of numerical method and
the value of the nonlinear coupling parameter for each curve are shown in the
legend. \label{fig:nlinAvLDV}}
\end{figure}

In the case of the LDV equation, we can proceed in a similar manner. Let us
denote by $\psi_M(g;t)=g\langle\lvert\overline{\nabla^2(\nabla
h)^2}\rvert\rangle$ the absolute value of the nonlinear term of the LDV equation
averaged over space and realizations for the numerical method $M$.  Then, for a
given value of $g$ used with the FD method, we can estimate a $\tilde g$ for the
PS method leading to a nonlinear term of the same order. This is
\begin{equation}\label{eq:gLDV}
  \tilde g=g
  \left\langle
  \frac{\psi_\text{FD}(g; t)}{\psi_\text{PS}(g; t)}
  \right\rangle_t.
\end{equation}
In the previous expression the angular brackets denote an average over the time
interval used in the simulation. As for the KPZ equation, the ratio
$\tilde g/g$ computed according to Eq. \eqref{eq:gLDV} depends slightly on both
the system size and $g$. For example, for a system
size of $L=128$ and a value of $g_\text{FD}=1.25$ used in the FD algorithm, we
find that a value of $g_\text{PS}\simeq 0.42$ gives rise to a nonlinear term of
the same order for the PS method (see Fig. \ref{fig:nlinAvLDV}). As occurs for
the KPZ equation, the nonlinear effects are stronger for the PS method.
%================ BEGIN FIGURE ==============
%======== END FIGURE ===========

We checked that the global dynamical exponents obtained with the FD and PS
methods are the same using values of $g$ according to \eqref{eq:gKPZ} and
\eqref{eq:gLDV}. The global interface width scales according to the
Family-Vicsek ansatz~\cite{Family}:
\begin{equation*}
  W(L,t)=\bigl\langle\,\overline{(h(x,t)-\bar h)^2}\,\bigr\rangle
  =t^{\alpha/z}\, f(L/t^{1/z}),
\end{equation*}
where the scaling function $f$ behaves as
\begin{equation*}
  f(u)\sim
  \begin{cases}
  u^\alpha\quad &\text{if $u\ll 1$},\\
  \text{const} & \text{if $u\gg 1$}.
  \end{cases}
\end{equation*}
The parameter $\alpha$ is the roughness exponent, $z$ is the dynamic exponent,
and the ratio $\beta=\alpha/z$ is the time exponent. In 1+1 dimensions the
critical exponents can be computed exactly~\cite{Kardar} and their values are
$\alpha=1/2$ and $z=3/2$, so that $\beta=1/3$. Using the FD with 
$g=2.5$, we found the exponents $\alpha\simeq 0.49$, $\beta\simeq 0.32$,
$z=\alpha/\beta\simeq 1.52$ for the FD method and with $g=1.2$ we obtain the
same values of the exponents with the PS method within error bars. For the LDV
equation the global exponents are known for arbitrary dimension. In 1+1
dimensions they are $\alpha=1$, $\beta=1/3$, and $z=3$. Taking a value of
$g=1.25$ we found the exact value of the exponents with two significant digits
by integrating the equation with the FD method, with system sizes ranging from
$L=16$ to $L=256$ and averaging the interfaces over 100 runs. On the other hand,
the PS method with $g=0.42$ provides the same exponents within error bars.

%========================================
\subsection{Stability of the algorithms}
%========================================

We tested the stability of the algorithms by measuring the probability $P(t)$
that the system exhibits a numerical overflow when starting from a flat
interface. This is measured from a large number of independent runs as the
frequency probability of getting a computer overflow at time $t$. This numerical
instability takes place when the height of the interface tends to grow
indefinitely. The probability of instability is a decreasing function of the
time step used in the simulations. 

\begin{figure}
\includegraphics[width=0.4\textwidth]{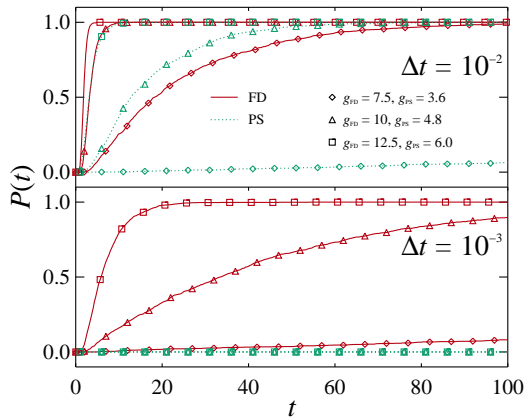}
\caption{Probability of instability for the 1D KPZ equation as a function of time.
The initial condition is a flat interface. Here $L=100$ and the probabilities
are computed using 2000 realizations. Curves for two values of the time step are
shown. We show results for the FD and PS numerical methods and some different
values of the nonlinear coupling parameter $g$ which are shown in the legend.
\label{fig:probFPE_KPZ}}
\end{figure}

In Fig.~\ref{fig:probFPE_KPZ} the probability of instability as a function of
time for the 1D KPZ equation is shown for several cases. We
show curves for two time steps $\Delta t=10^{-2}$ and $\Delta t=10^{-3}$. The
system size is $L=100$ and the probabilities are computed over 2000 samples. The
values of $g$ were chosen in such a way that the nonlinear effects for the two
methods were of the same order. For the KPZ equation this is achieved when
$g_\text{PS}\simeq0.48 g_\text{FD}$. In all cases we found the probabilities for
the PS method to be smaller than those of the FD method for a given time step.
For example, for a time step of $\Delta t=10^{-3}$, we can see in the bottom
graphic of Fig.~\ref{fig:probFPE_KPZ} that the PS method is stable (that is, the
probability of instability is equal to zero) in the time interval $[0,100]$ for
values of $g=3.6$, $4.8$, and  $6.0$, whereas the FD becomes unstable at very
short times. 

\begin{figure}
\includegraphics[width=0.4\textwidth]{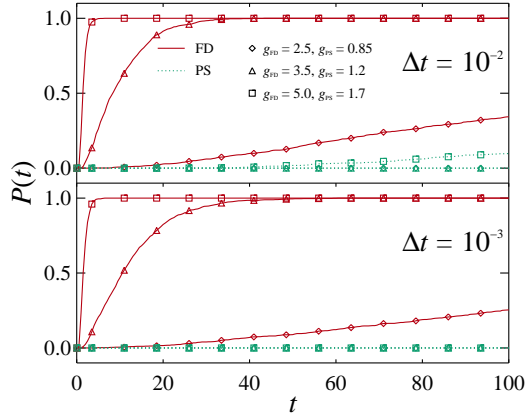}
\caption{Probability of instability for the 1D LDV equation as a function of time.
The initial condition is a flat interface. Here $L=100$ and the probabilities
are computed using 2000 realizations. Curves for two values of the time step are
shown. We show results for the FD and PS numerical methods and some different
values of the nonlinear coupling parameter $g$ which are shown in the legend.
\label{fig:probFPE_LDV}}
\end{figure}

In Fig.~\ref{fig:probFPE_LDV} we show the probability of instability as a
function of time for the 1D LDV equation. In this case we took
$g_{\text{PS}}=0.34g_\text{FD}$ to match the nonlinear effects for both methods.
In much the same way as for the KPZ equation we see that for a given time step
the PS method is the most stable. This is also observed for other time steps 
ranging from $10^{-5}$ to $10^{-2}$. We then conclude that the PS method is the 
most stable when the intensity of the nonlinear terms are of equivalent
magnitude. This means that, under the same conditions, the PS method
allows to follow the temporal evolution of the system up to larger times.
%================ BEGIN FIGURE ==============
%======== END FIGURE ===========

%================ BEGIN FIGURE ==============
%======== END FIGURE ===========

%=========================
\subsection{KPZ equation}
%=========================

There are some exact results of the continuum KPZ model that we can used to test
the numerical methods. First, the steady state probability distribution of the
heights is known exactly~\cite{Barabasi,Krug-rev}. In terms of the slopes,
$m(x)=\partial_x h(x)$, it is known that
\begin{equation*}
  \mathcal P(m)\sim\exp\left[-\int dx\,m(x)^2\right].
\end{equation*}
This expression can be written approximately as:
\begin{equation*}
  \mathcal P(m)\sim\exp\left(-
    \sum_{i=0}^n\Delta x\, m_i^2\right)=
  \exp\left(-{\Delta x}\,m^2\right).
\end{equation*}
We have used the central limit theorem to identify $\sum_{i=0}^n m_i^2$ with
$m^2$. Here $m$ represents the slope of the field in the steady state at any
point of the lattice.The normalized expression of the probability is:
\begin{equation*}
  \mathcal P(m)=\pi^{-1/2}\,e^{-m^2}.
\end{equation*}
With this probability distribution, we can find the several moments of
the slope $m$: $\langle{m^2}\rangle=1/2$, $\langle{m^4}\rangle=3/4$, 
$\langle{m^6}\rangle=15/8$.

For the discretized model and when the LS discretization is used, the
steady state distribution probability is found to be~\cite{Lam}
\begin{equation}\label{eq:P_hi}
  P[h_i]\sim\exp\left[-\sum_{i=0}^N(h_{i+1}-h_i)^2\right].
\end{equation}

It is worth mentioning here a caveat concerning~\eqref{eq:P_hi}. One can see
that, in order to reproduce numerically the slope distribution with the FD
method, slopes must be computed with the forward (or backward) operator
(Eq.~\eqref{eq:opf}), so that  $m_i=h'_i=h_{i+1}-h_i$. We have checked that if
the symmetric rule to compute the derivatives ($m_i=(h_{i+1}-h_{i-1})/2$) is
used the width of the slope probability distribution is far from unity, which is
the exact value. When forward or backward derivatives are used, however, the
correct value is recovered. Remarkably, the PS method provides the proper result
in a natural way.

The global interface width in the steady state is also known
exactly~\cite{Krug_Plisch},
\begin{equation}\label{eq:GwidthSteady}
  W(L)=\sqrt{\dfrac{1}{24}}\,L^{1/2},\quad t\rightarrow\infty,
\end{equation}
and it is independent of the nonlinear coupling parameter $g$. In
reference~\cite{Lam} it is shown that a FD method with conventional
discretization for the nonlinear term, Eq.~\eqref{eq:usualDiscKPZ}, provides
steady state interfaces whose global width is of the
form~\eqref{eq:GwidthSteady} but with a prefactor of $L^{1/2}$ significantly
smaller than the predicted value $24^{-1/2}$. It has been argued~\cite{Lam} that
with the improved discretization \eqref{eq:LSDiscKPZ} the correct value for the
prefactor is recovered. A plot of $\phi(L)=\sqrt{24/L}\,W(L)$ versus $L^{-1}$
was presented in references~\cite{Giada2,Giacometti2}, showing that $\phi(L)$ is
unity within error bars for both the PS method and the FD method with the
discretization \eqref{eq:LSDiscKPZ}, although the dispersion of the data is
larger for the FD method. We have also carried out a similar study comparing FD
and PS methods. We checked that the curve $W(L)$ {\it vs.} $L$ can be fitted to
a function of the form $B\,L^{1/2}$, where $B=0.182\pm 0.002$ for the FD method
with the usual discretization \eqref{eq:usualDiscKPZ}. As expected, this value
is clearly smaller than the nominal value $B_0=24^{-1/2} \simeq 0.204$. This
observation is in agreement with that of reference~\cite{Lam}. On the other
hand, for both PS and FD method with LS improved discretization
\eqref{eq:LSDiscKPZ} we obtain the same value (indistinguishable up to the third
digit) $B=0.196\pm 0.003$, a value very close to $B_0$ indeed. Therefore, with
the PS method we obtain in a natural way the result predicted by the continuum
model for the steady state global interface width. For the FD method, on the
contrary, we must use an {\em ad hoc} discretization of the nonlinear terms to
achieve the same results.

%=========================
\subsection{LDV equation}
%=========================

\begin{figure}
\includegraphics[width=\textwidth]{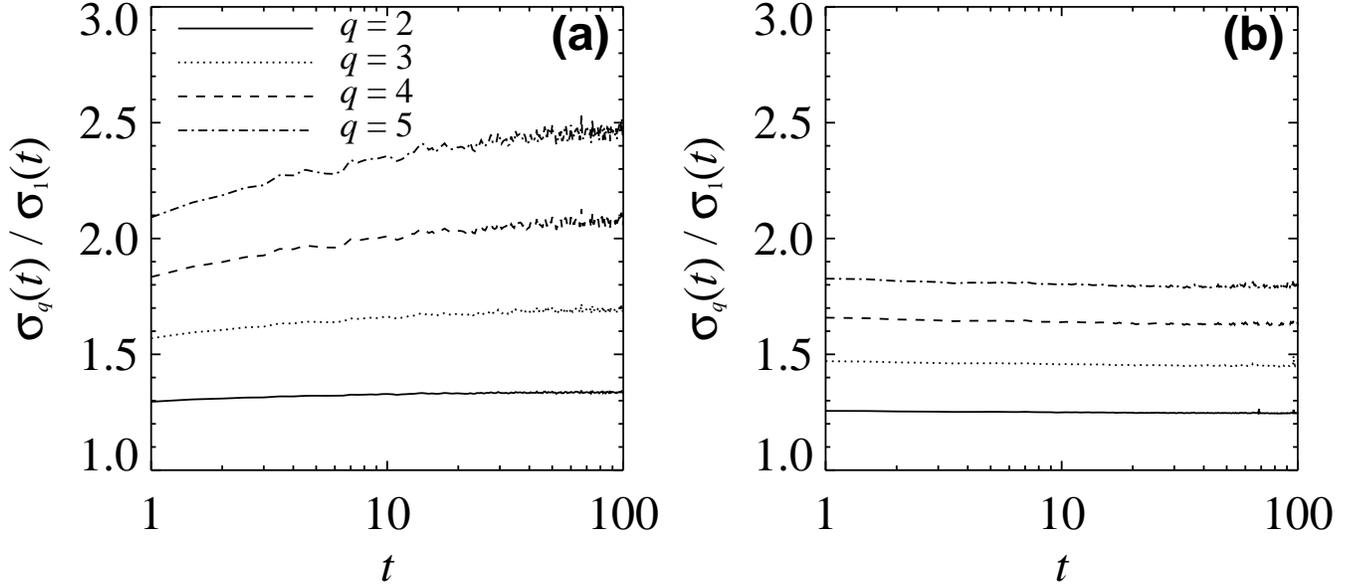}
\caption{Ratios $\sigma_q/\sigma_1$, $q=2,3,4,5$,  as a function of time, of the
moments of the nearest neighbor height difference for the 1D LDV equation
integrated with the FD [Fig. (a)] and PS [Fig. (b)] numerical methods. Here
$L=100$, $g_\text{FD}=2$, $g_\text{PS}=1.5$, and the averages are taken over
300 realizations. \label{fig:multiLDV}}
\end{figure}

We have investigated the influence of the numerical method on the multiscaling
behaviour of the LDV equation. The multiscaling can be detected by looking at
the moments of the nearest neighbour height difference. We define~\cite{Krug}
\begin{equation*}
  \sigma_q(t)=\langle\lvert h_{i+1}(t)-h_i(t)\rvert^q\rangle^{1/q},
\end{equation*}
where the average is taken over each site of the system and over the different
realizations of the noise. In systems that exhibit anomalous scaling, as the
case of the LDV equation for intermediate times (see note~\cite{note}), the
moments $\sigma_q(t)$ are expected to grow as follows
\begin{equation*}
  \sigma_q(t)\sim t^{\alpha_q/z},\quad t\ll L^z,
\end{equation*}
where $L$ is the system size and $z$ is the dynamical exponent. When the moments
scale in a different way, that is, when the \mbox{$\alpha_q$'s} depend on $q$,
the system is said to show multiscaling. As done in~\cite{Dasgupta,Dasgupta2},
we monitorize the multiscaling by looking at the ratios
$\sigma_q(t)/\sigma_1(t)$, $q\geq 2$. On the other hand, the multiscaling can
also be studied by measuring the height difference correlation
function~\cite{Krug, Dasgupta, Dasgupta2}. In Fig. \ref{fig:multiLDV}a we show
the ratios $\sigma_q(t)/\sigma_1(t)$ with $q=2,3,4,5$ for the 1D LDV equation
integrated with the FD scheme. Parameter values are $L=1000$, $g=2$, and the
averages are taken over 100 realizations. This is in fact a reproduction of Fig.
12 of reference~\cite{Dasgupta2}. As can be seen in the picture, the greater the
 $q$, the faster the growth of $\sigma_q/\sigma_1$ with time. This behaviour
implies the existence of multiscaling. For times greater than 100, the evolution
of the system cannot be followed due to the presence of numerical instabilities.
The authors of \cite{Dasgupta2} claimed that this behavior is related to an
instability of the discretized LDV equation against the growth of isolated
pillars, which are just height profiles such that the field $h$ is positive at a
certain point while being zero otherwise. For a given a value of the parameter
$g$, there exists a critical value $h_c$ of the pillar height beyond which the
pillar grows with a certain probability. It is found that this critical height
goes as $h_c\sim g^{-1}$. On the other hand, the effect of the magnitude of the
time step on this instability seems to be very small. For this reason, it is
further argued in \cite{Dasgupta2} that this instability is {\em not} a
numerical
artifact due to the use of a not small enough integration time step.
As we show in the following our results disagree with this interpretation.
%================ BEGIN FIGURE ==============
%======== END FIGURE ===========

Interestingly, we find that the numerical integration of the LDV equation with
the PS method has a significant impact on the observed scaling properties. The
instability discussed above, which is present with any FD method, is no longer
present, at least in the wide range of couplings we studied. Specifically, a
pillar like initial condition of the form
\begin{equation}\label{fig:iniPillar}
  h_j=\begin{cases}
    h_0\quad &\text{if $j=\frac{N-1}{2}$},\\
    0 & \text{otherwise.}
  \end{cases},\quad \text{with $0\leqs j\leqs N-1$, $N$ odd, $h_0>0$}
\end{equation}
never grows with the PS method. In the absence of noise, the temporal derivative
of the field \eqref{fig:iniPillar} is always negative. A straightforward
calculation leads to
\begin{equation*}
  \partial_t h=-\nabla^4h+g\nabla^2(\nabla h)^2=
  -\frac{h_0\pi^4}{15L^4}(N^2-1)(3N^2-7)\approx 
  -\frac{h_0\pi^4}{5}<0,\ L=N\gg 1.
\end{equation*}
Therefore, a pillar or the form \eqref{fig:iniPillar} always tends to shrink in
the deterministic case. Other structures like double pillars, however, might
grow in time when their size exceeds a certain value.

In Fig. \ref{fig:multiLDV}b we show the ratios $\sigma_q(t)/\sigma_1(t)$, 
$q=2,3,4,5$,  for the 1D LDV equation integrated with the PS scheme. In this
case we use a value of $g=1.5$  for which the nonlinearities are considerably
stronger than for the FD method with $g=2$. As it can be observed, the curves do
not grow in time, which means that there is no multiscaling. For other values of
$g$ and other system sizes similar results were obtained. It is worth
mentioning, however, that the instability does show up when the aliasing is {\em
not} removed. In this case an isolated pillar may grow when its height is larger
than a critical value, in much the same way as with the FD method. So, it is
strongly recommended to remove aliasing effects when applying a PS method to
correctly describe the continuum physics. Note that  the aliasing does not
actually affect global properties of the system such as multiscaling, the value
of the critical exponents, etc. Indeed, we have also carried out a study of the
PS method without  removing the aliasing because in this case a similar
instability of the FD method is present. We have not observed multiscaling in
this case either. We then conclude that the multiscaling does not seem to be
related to this instability for the PS method.

This analysis leads to the main conclusion of our paper that the existence of
multiscaling may depend on the numerical scheme used to integrate the growth
model. We argue that the instability (and associated multiscaling behavior) is
intrinsic to the numerical integration scheme rather than to the discretization
itself, in contrast with the conclusions of Ref.\ \cite{Dasgupta, Dasgupta2}.
Our results clearly show that the instability previously found in FD
discretizations has to be seen as spurious and inherent to the discretization
scheme used. PS integration methods do not show any trace of either instability
(when aliasing is properly removed) or multiscaling, representing much more
accurately the dynamics and statistics of the continuum problem. We conclude
that a PS method should be preferred for surface growth equations with anomalous
scaling, because next-neighbor height differences in this case can grow very
large.

%%%%%%%%%%%%%%%%%%%%%
\section{Conclusions}
%%%%%%%%%%%%%%%%%%%%%

We have shown that the choice of the numerical method used to integrate certain
stochastic models of surface growth may be of paramount importance in the study
of some physical properties of the system. We have compared a standard finite
difference method with a pseudospectral numerical scheme in the integration of
the KPZ and LDV growth models in 1+1 dimensions. As the FD method underestimates
the nonlinear effects with respect to the PS method, the nonlinear coupling
parameter were tuned up so that the nonlinear terms were of the same order for
both numerical methods on average. The global critical exponents, obtained from
the global interface width are the same for the two numerical methods. 

With regard to the KPZ equation there are some exact results available derived
from the continuum model. The expression for the global width of the interface
is known in the saturation regime. With the FD method and a standard
discretization for the nonlinear term, the amplitude of the width of the numerical
interfaces is smaller than that of the continuum.  With the spectral
approach, on the contrary, numerical results are very close to the predicted
value. Nevertheless, it is possible with the finite differences scheme to get
close to the continuum model prediction for the width of steady state
interfaces, but at the expense of using more sophisticated discretizations.

We have tested the stability of the algorithms by measuring for different time
steps the probability of the system to undergo a floating point instability
evolving from a flat interface. This instability is related to a numerical
overflow in the surface height data. The PS method proved to be the most stable
in all the cases we have studied for both models. In the same way, with the PS
method it is possible to follow the temporal evolution of the system for longer
times than with the FD method.

The LDV equation exhibits anomalous scaling at intermediate times, so that
according to~\cite{Lopez99} (but see also~\cite{note}) the average slope of the
field is expected to grow in time. Any FD method leads to a numerical
instability against the growth of an isolated pillar appears. This instability
has been claimed~\cite{Dasgupta,Dasgupta2} to be the reason why approximate
multiscaling is observed in the numerical simulations, although multiscaling is
not present in the continuum equation. More importantly, this multiscaling has
been interpreted as a real physical effect, which could explain the multiscaling
behavior of surface fluctuations observed in atomistic models believed to belong
to the LDV universality class. However, our results show that this
interpretation is misleading. We have shown that surface multiscaling is not
observed with the PS method, regardless of the temporal evolution of isolated
pillars. Therefore, surface multiscaling does not seem to be related to this
instability for the PS method nor represent any intrinsic physics of the LDV
equation as such. In this respect, due to the fact that discrete models can be
mapped into continuum equations, in particular the LDV
equation~\cite{Vvedensky}, our results indicate that multiscaling behavior
observed in such systems could be an artifact of the discretization of the
dynamics and, consequently, the are not intrinsic to the physical system they
are trying to modelize.

\begin{acknowledgments}
This work is supported by the DGI of the Ministerio de Educaci{\'o}n y Ciencia 
(Spain) through Grant Nos. FIS2006-12253-C06-04 and FIS2006-12253-C06-06.
\end{acknowledgments}


\begin{thebibliography}{10}
\bibitem{note} This anomalous exponent should be interpreted as an
effective exponent according to recent theoretical results on the neccesary conditions for
anomalous roughening in local growth models~\cite{Lopez05}. Nevertheless, the effect at
intermediate times (before the truly asymptotic regime referred to in~\cite{Lopez05}) is
to produce a power-law growth of the local slopes and therefore instabilities of the
finite-differences integration scheme as described.
\end{thebibliography}
\end{document}